\documentclass[fleqn,12pt,twoside]{article}
\usepackage[headings]{espcrc1}

\readRCS
$Id: sewm.tex,v 1.3 2006/06/23 07:38:00 pbialas Exp pbialas $
\usepackage{graphicx}
\usepackage[figuresright]{rotating}

\newcommand{\AmS}{{\protect\the\textfont2
  A\kern-.1667em\lower.5ex\hbox{M}\kern-.125emS}}

\title{SU(3) gauge theory at finite temperature in 2 + 1 dimensions}

\author{P. Bialas,\address[MCSD]{Institute for Physics, Jagellonian 
University, \\ 
       ul. Reymonta 4, 30 059 Krakow, Poland}
        L. Daniel,\addressmark 
        A. Morel\address{Service de Physique Theorique, CEA Saclay \\
        F-91191 Gif-sur-Yvette Cedex, France}
        and
        B. Petersson\address{Fakulty for Physics, University of Bielefeld,\\
  P.O.Box 10 01 31, D-33501 Bielefeld, Germany}}

\runtitle{SU(3) gauge theory at finite temperature in 2 + 1 dimensions}
\runauthor{B. Petersson}

\begin{document}

\maketitle

\begin{abstract}
In this article we will discuss numerical results on screening 
masses and thermodynamic quantities in 2 + 1 dimensional SU(3) gauge 
theory. We will also compare them to perturbation theory and the dimensionally 
reduced model.  
\end{abstract}


\bigskip

The SU(3) gauge theory in 2+1 dimensions is simple enough from a
numerical point of view, so that it is possible with present computers
to make continuum extrapolations with controlled systematic errors.
Of course, there are some obvious differences between SU(3) gauge
theory in 2 + 1 and 3 + 1 dimensions. In 2 + 1 dimensions the coupling
constant $g^2$ has the dimension of a mass, and the theory is
superrenormalizable.  The tree level potential between heavy quarks is
already logarithmically confining: $V(r) \sim \log r$.  There are,
however, many similarities. One may introduce a dimensionless
``running'' coupling constant $g_3 (l)$ by the definition $g^2_3 (l)
\equiv l g^2$ where $l$ is a length scale. Then $g^2_3
(l) \rightarrow 0$ for $l \rightarrow 0$ and to infinity for $l
\rightarrow \infty$.  This is somewhat analogous to the
logarithmically running coupling constant in 3 + 1 dimensional SU(3)
gauge theory.  In 2 + 1 dimensions the coupling constant $g^2$ sets
the scale, and $m_i = c_i g^2$, where $c_i$'s are numerical constants.
From Monte Carlo simulations one knows some further properties: There
is a linearly rising non-perturbative potential $V(r) \simeq \sigma_ 0
r$ for $r$ large\cite{teper,lego}. There is a second order phase
transition at \hbox{$T_c = 0.55(1) g^2$ ,} with the critical indices
of the 2d 3states Potts model\cite{lego}.  Furthermore, the glue ball
masses $m_{GB}$ are much bigger than $T_c, \, m_{GB} \ge 4.4 T_c$
\cite{teper}. This is all qualitatively similar to 3 + 1 dimensions,
where, however, the transition is weakly first order. In the gluon
plasma phase $T > T_c$, one should be able to use perturbation
theory.  The relevant dimensionless coupling in 2 + 1 dimensions  is
\begin{equation}
g^2_3 (T) \equiv \frac{g^2}{T} \mathop{\longrightarrow}\limits^{T\rightarrow
\infty} 0 . 
\end{equation}   
There are, however, infrared divergences, which are even more serious
than in 3 + 1 dimensions. For the screening mass (rsp. the pressure)
they appear already at order $g^2$, i.e. at one (resp. two) loop(s).
The infrared divergences can be tamed through resummations, e.g.
through the selfconsistent perturbation theory (SCPT) introduced by
D'Hoker\cite{hoker}.  It works in the following way. Choose the class
of static gauges $\partial_0 A_0 (x) = 0$ where the Polyakov loop is
purely static,
\begin{equation}
L (\bar x ) = T r \, e^{i \frac{g}{T} A_0 (\bar x)}  .
\end{equation}
Add and subtract an explicit mass term in the static sector,
\begin{equation}
\frac{1}{2} m^2_g \, T r (A_0 (\bar x))^2 , 
\end{equation} 
which is invariant inside this class of gauges, and perform the
perturbative expansion in the theory with $A_0$ massive. In fact,
there are no further infrared divergencies in 2+1 dimensions\cite{hoker}.

The second method, which is semianalytic, is the dimensional reduction
from \hbox{2 + 1} to 2 dimensions\cite{pb1}. Again we choose the class of
static gauges. We then integrate perturbatively over the non-static
modes. This integration is by construction infrared finite. The result
is an effective 2-dimensional adjoint Higgs model for the static modes
with the action
\begin{eqnarray}\label{eq:higgs}
S = & & \int \, d^2 x T r \left\{ \frac{1}{2} F_{ij}^2 + (D_i \phi )^2 
- \frac{3 g^2 T}{2\pi} \left[ \frac{5}{2} \log 2 + 1 - \log (a T) 
\right] \phi^2 
+ \frac{g^4}{32 \pi} \phi^4 \right\} 
\end{eqnarray}
where $D_i$ is a covariant derivative and where the coupling constants
are derived from a one loop integration over the static modes. This
action is systematic in $\frac{g^2}{T}$, e.g.  the term $\phi^6$ is
multiplied by a constant proportional to $\frac{g^6}{T}$, and is
neglected at high $T$.
\begin{figure}[t] 
\includegraphics*[width=9.0cm]{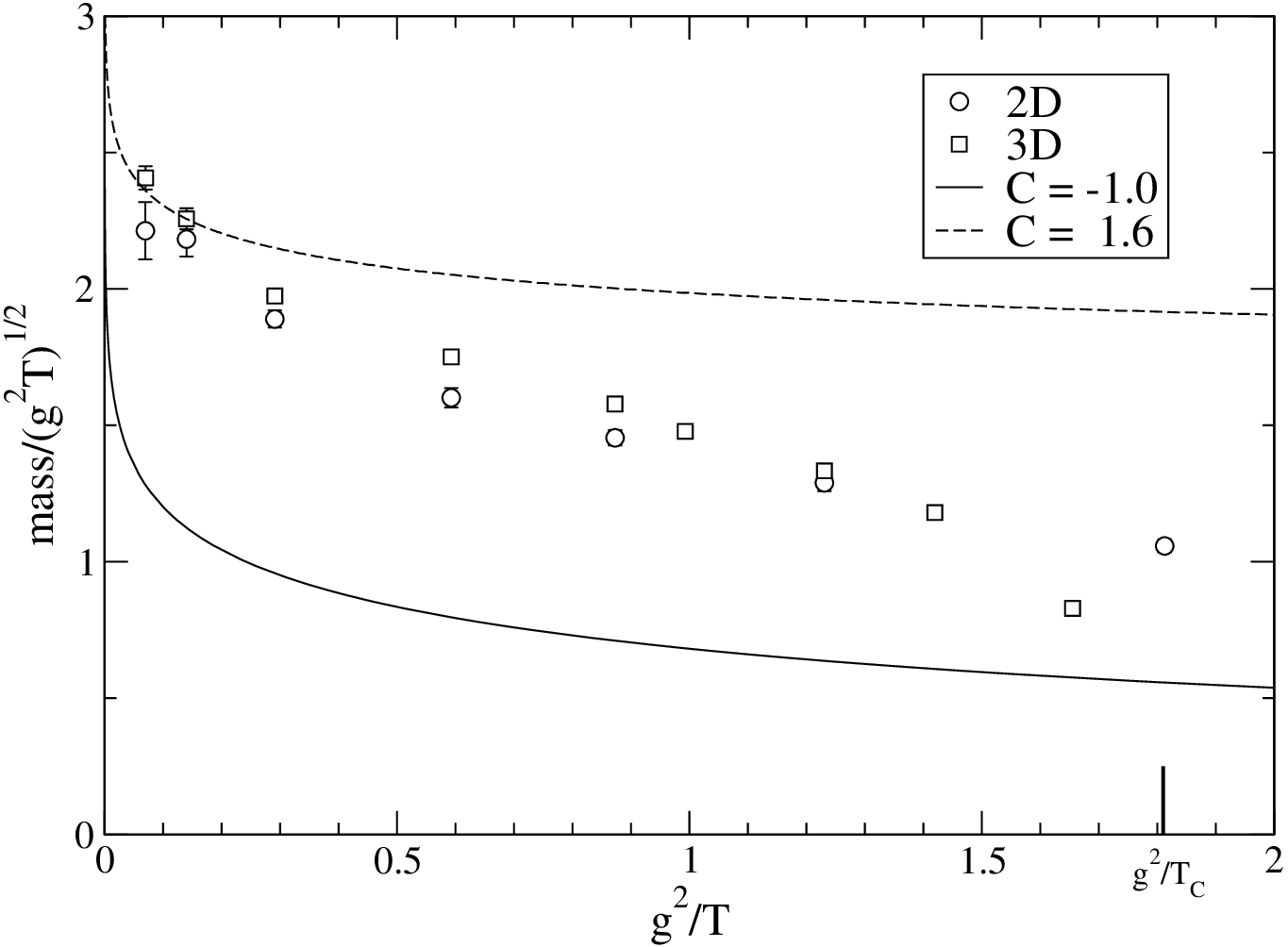}
\vspace{-10mm}
\caption{\label{fig:hooker} Screening masses from the  3D SU(3) theory on
lattices with temporal extent $N_\tau = 4$ compared to 
the predictions of SCPT(continuous lines) and to the
2D reduced model.}  
\end{figure}

The two-dimensional adjoint Higgs model has not been solved
analytically.  We solve it non-perturbatively by a Monte Carlo lattice
simulation.

We define a screening mass for $T > T_c$ by
\begin{equation}\label{eq:ms}
\langle Re \left( L (0) L^\dagger(\vec x )\right) \rangle \simeq
\langle L \rangle^2 + 
\frac{A}{\sqrt{m |\vec x |}}e^{-m | \vec x |}
\end{equation}
In lowest order perturbation theory $m = 2 m_g$ for the correlations of 
$Re L (\bar x)$ and $m = 3 m_g$ for the correlations of $Im \, L(\bar x)$.
In SCPT one has\cite{hoker} 
\begin{equation}\label{eq:scpt}
\frac{m^2_g}{g^2 T} = \frac{3}{2\pi} \left( \log \frac{T}{m_g} + C 
\right) + {\cal O} \left( 1 / \log \frac{T}{m_g} \right),\quad C=-1.0.
\end{equation} 
In Fig.~\ref{fig:hooker}, we show $2 m_g$ compared to the screening
mass in the 2 + 1 dimensional SU(3) gauge theory. We have used the
formula $T/T_c=(\beta-1.5)/3.3 N_\tau$, derived from the condition
$T_c/g^2=0.55$ for all $N_\tau$, and the values $\beta_c(N_\tau)$
from\cite{lego}. Solving Eq.~(\ref{eq:scpt}) where $m_g/T$ has been replaced by $(1-\exp(-m_g/T))$,  
we can get agreement
only for  $T/g^2>6$ ($T/T_c>12$), and this only by arbitrarily
choosing $C=1.6$.

The dimensionally reduced model is in good agreement with the full
theory already for $T>1.5 T_c$ as can be seen in
Fig.~\ref{fig:hooker}.  A further investigation showed that the two
exchanged states in the reduced model are simple poles, not 2 gluon
and 3 gluon cuts respectively\cite{pb2}.

For the thermodynamics we use the lattice integral method introduced
in \cite{engels}. 
\begin{figure} 
\includegraphics[width=9.0cm]{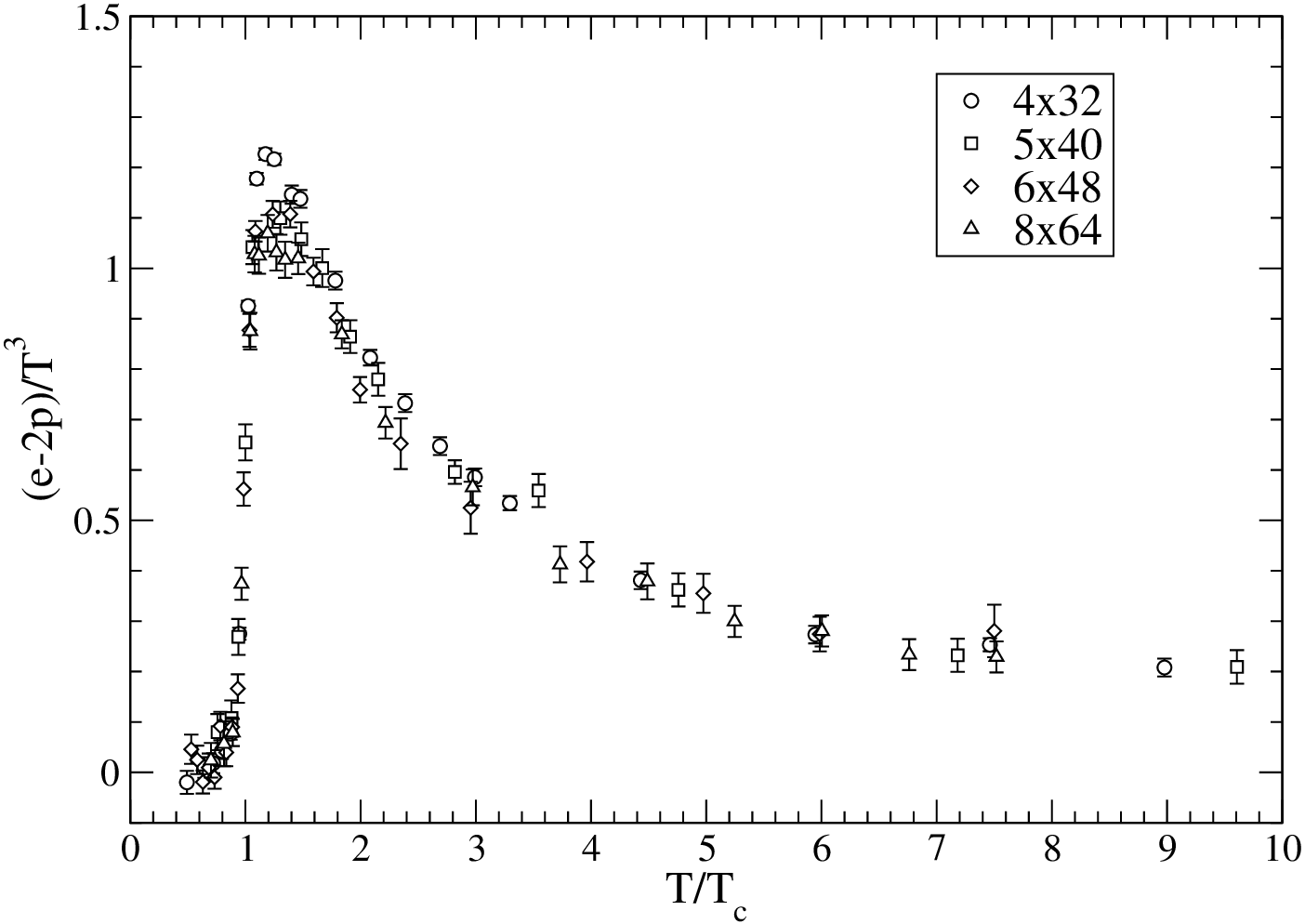} 
\vspace{-10mm}
\caption{\label{fig:trace}Trace anomaly $(\epsilon - 2 p) / T^3$,
  on  lattices of various sizes.}
\end{figure}
The trace anomaly,
$(\epsilon - 2 p) / T^3$, is presented in
Fig.~\ref{fig:trace}.  It is qualitatively similar to the result in 
\hbox{3 + 1} dimensions. 

Since the trace anomaly is
zero for free massless gluons, we expect in
perturbation theory that 
\begin{equation}\label{eq:scal}
\frac{\epsilon - 2 p}{T^3} \simeq  \frac{g^2}{T} f \left( \log \frac{T}
{g^2} \right)  .
\end{equation}
\begin{figure}[h] 
\includegraphics*[width=9.0cm]{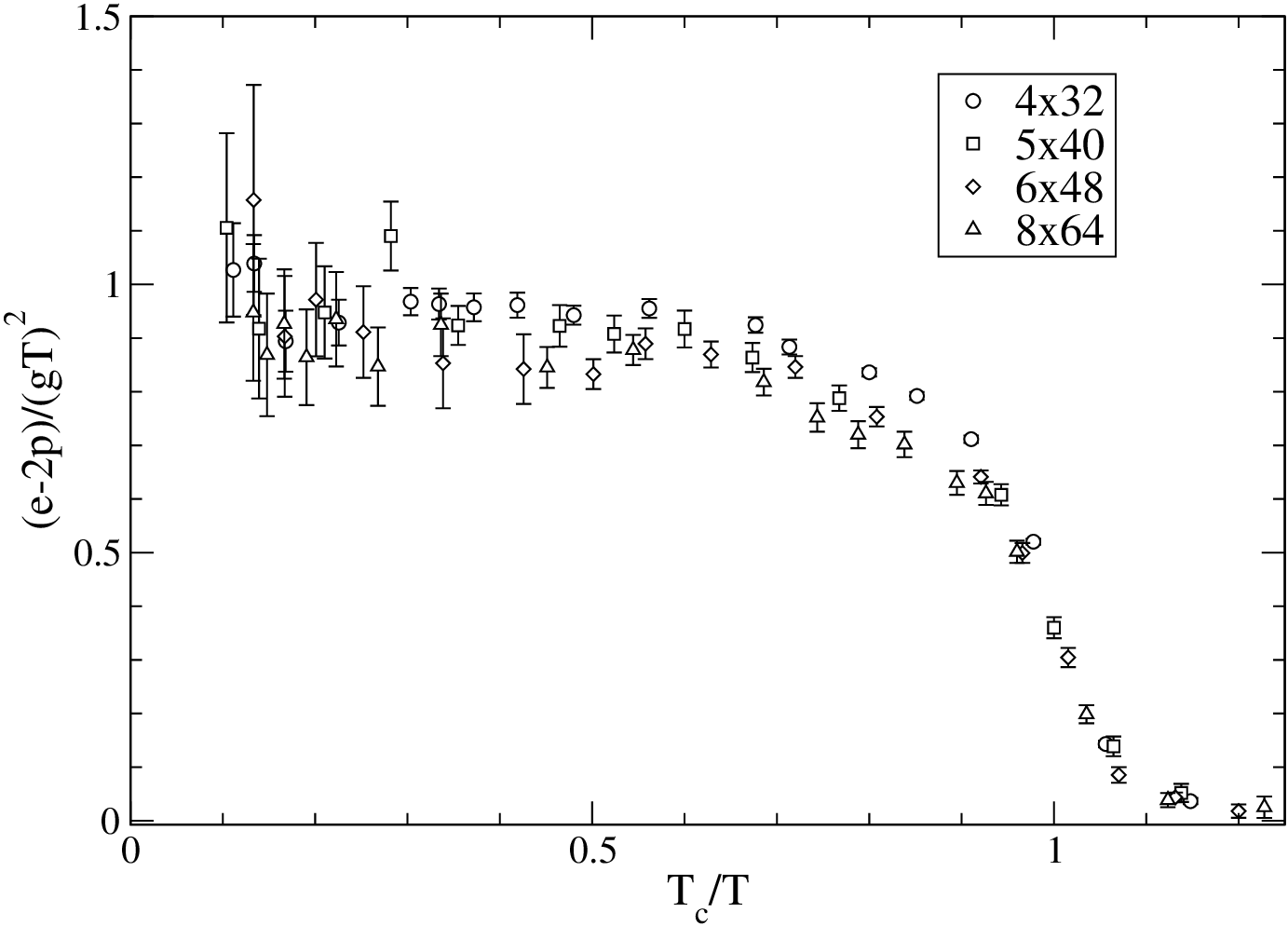}
\vspace{-10mm}
\caption{\label{fig:scal}Check of the scaling formula (\ref{eq:scal}).}
\end{figure}
In Fig.~\ref{fig:scal} we present $(\epsilon - 2 p) / T^2 g^2$ as a
function of $T_c / T$. One observes that this quantity is
slowly varying at high temperature. A comparison with
perturbation theory and dimensional reduction is in progress. 
\begin{figure}[h] 
\includegraphics*[width=9.0cm]{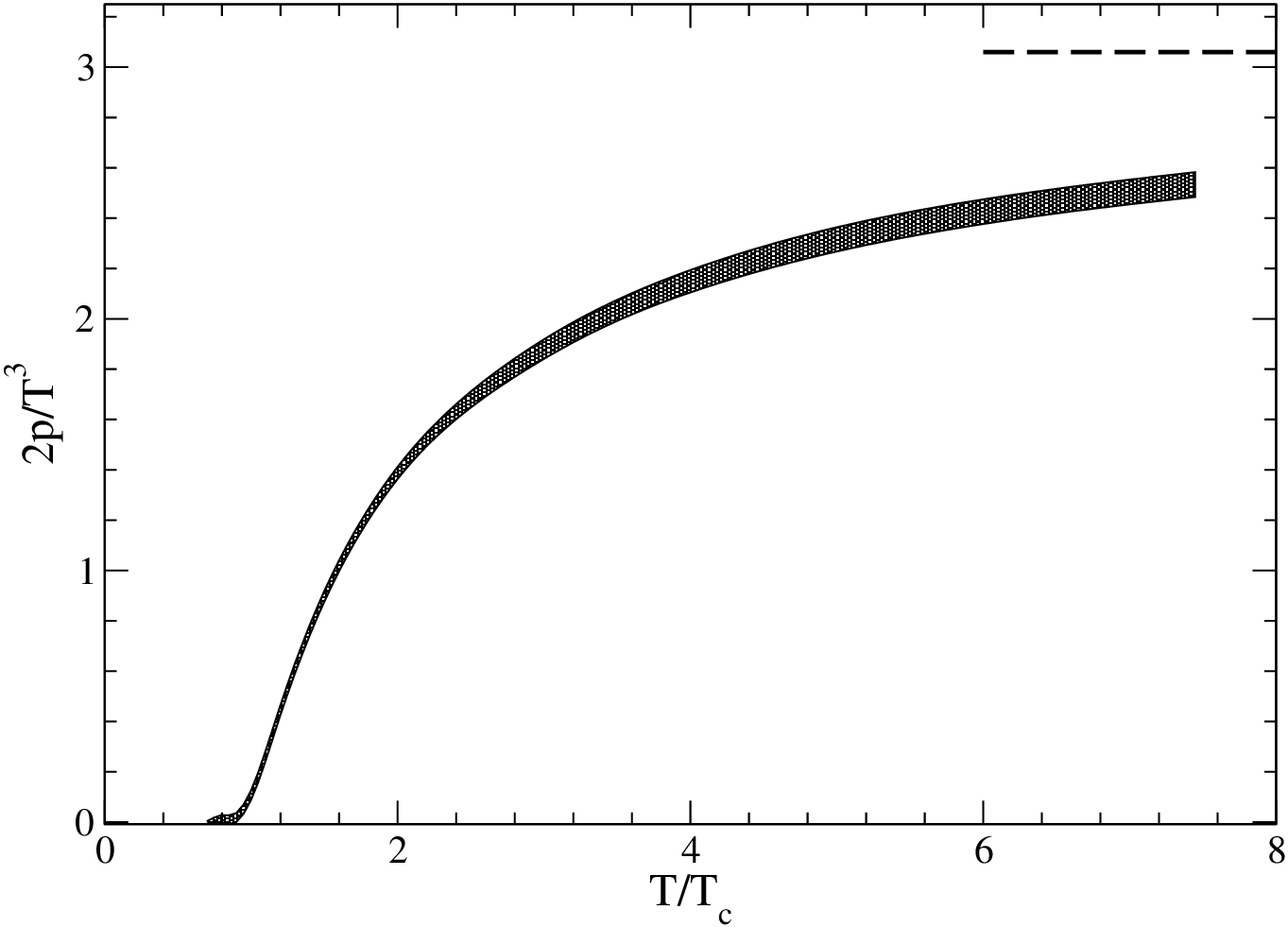}
\vspace{-10mm}
\caption{The pressure extrapolated to the infinite
$N_\tau$ limit from the lattices of size $4\times32^2$, $6\times48^2$
and $8\times64^2$ using a linear fit in $1/N^2_\tau$. The dashed line
shows the Stefan-Boltzman limit.}
\end{figure}

In Fig. 4, we present the extrapolation of the pressure
to the continuum. 

This work is supported through ENRAGE (European Network on Random
Geometry), a Marie Curie Research Training Network supported by the
European Community's Sixth Framework Programme, network contract
MRTN-CT-2004-005616.

\end{document}